\documentclass{jetpl2} 

\usepackage{amsmath}
\usepackage{amssymb}
\usepackage{graphics}
\usepackage{epsfig}
\usepackage[dvipdfm]{hyperref}
\twocolumn

\lat

\DeclareMathOperator{\tr}{tr}
\DeclareMathOperator{\Df}{{\mathcal{ D}}}      
\DeclareMathOperator{\lagr}{\mathcal{ L}}

\title{On formation of equation of state of evolving quantum field}

\rtitle{On formation of equation of state of evolving quantum field}

\sodtitle{On formation of equation of state of evolving quantum field}

\author{A.\,V.\,Leonidov$^{*,**,***}$\thanks[1]{Also at ITEP, Moscow, Russia}, 
A.\,A.\,Radovskaya$^{*}$\thanks[2]{e-mail: aradovsk@cern.ch}}

\rauthor{A.\,V.\,Leonidov, A.\,A.\,Radovskaya}

\sodauthor{Leonidov, Radovskaya}

\address{$^*$ P.N. Lebedev Physical Institute, 119991 Moscow, Russia\\
$^{**}$  Moscow Engineering Physics Institute, Moscow, Russia\\
$^{***}$  Moscow Institute of Physics and Technology, Moscow, Russia}

\begin{document}

\abstract{Stylized model of evolution of matter created in ultrarelativistic heavy ion collisions is considered.
Systematic procedure of computing quantum corrections in the framework of Keldysh formalism is formulated.
Analytical expressions for formation of equations of state taking into account leading quantum corrections are worked out,
complete description of subleading corrections and analytical expressions for some of them are presented. }


 \maketitle
 
 \section*{introduction}
 
 Quantitative understanding of physics of the early stages of ultrarelativistic heavy ion collisions remains, despite strong efforts, an outstandingly difficult problem. One of the most important issues is a possibility of applying hydrodynamic description that fits many observable quantities, see e.g. the recent reviews \cite{therm1,therm2,therm3}. For hydro description to be valid the system should become sufficiently equilibrated. In particular, a one-to-one relation between energy and pressure providing a well defined equation of state is required. The problem of formation of equation of state was analyzed, at an example of scalar field theory, in \cite{initial1,initial2,initial3}. The analysis was based on the fact that summation of leading quantum corrections can be cast in the form of integration over initial conditions for classical trajectories with the weight given by the Wigner function, see \cite{initial0} and, in different contexts, \cite{cosm,method,chem1,chem2}.   
 
The aim of this letter is to introduce a systematic formalism based on Keldysh technique \cite{Keldysh} allowing to compute subleading corrections to temporal evolution of observables. In particular, using the model of \cite{initial1}, we shall provide an analytical description of pressure relaxation in the leading approximation in quantum corrections as well as explicit equations for next-to-leading order corrections.

\section*{general formalism}

Let us consider temporal evolution of the observable $F(\hat\varphi)$ in the time interval $[t_0,t_1]$. The expectation value of $F(\hat\varphi)$ at the moment $t_1$ reads
\begin{gather}
  \langle F(\hat\varphi) \rangle_{t_1} =\tr (F(\hat\varphi) \hat\rho(t_1))=
   \int d\xi \int d\xi_1 \int d\xi_2  F(\xi) \nonumber\\
  \times \langle\xi|U(t_1,t_0)|\xi_1\rangle  \langle\xi_1|\hat\rho(t_0)|\xi_2\rangle
\langle\xi_2| \hat U(t_0,t_1)|\xi\rangle, \label{genev}
\end{gather}
where evolution of the density matrix $\hat\rho(t)$ is governed by the evolution operator $\hat U(t,t_0)$ 
\begin{equation}
 \hat\rho(t)=\hat U(t,t_0) \hat\rho(t_0) \hat U(t_0,t),
\end{equation}
and we have defined $\hat\varphi|\xi\rangle =\xi|\xi\rangle$.

\begin{figure}[h]
    \includegraphics[width=0.45\textwidth]{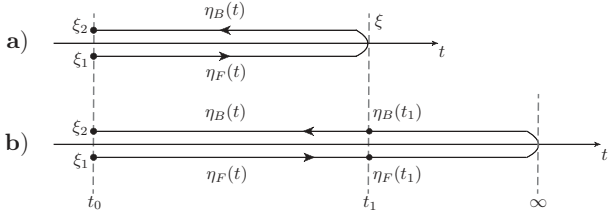}
     \caption{Fig.1:    a) Keldysh contour, b) extended  Keldysh contour}
      \label{fig1}
\end{figure}

The matrix elements of the evolution operator in Eq.~(\ref{genev}) for forward and backward time evolution are conveniently written in terms of the fields $\eta_{B,F}$ as
   \begin{equation}
        \langle\xi|\hat U(t_1,t_0)|\xi_1\rangle = \int\limits_{\eta_F(t_0)=\xi_1}^{\eta_F(t_1)
  =\xi} \Df \eta_F(t) e^{i S[\eta_F]} \label{forev}
  \end{equation}
  and 
  \begin{equation}
       \langle\xi_2|\hat U(t_0,t_1)|\xi\rangle = \int\limits_{\eta_B(t_0)=\xi_2}^{\eta_B(t_1)
=\xi} \Df \eta_B(t) e^{-i S[\eta_B]}, \label{backev}
\end{equation}
where
\begin{equation}
      S[\eta]=\int\limits_{t_0}^{t_1}\lagr (\eta,\partial_{t}\eta) dt'.
  \end{equation}
Using Eqs.~(\ref{forev},\ref{backev}) one can rewrite (\ref{genev}) in the following form:
  \begin{multline}
    \langle F(\hat\varphi)\rangle_{t_1} =\int d\xi \int d\xi_1 \int d\xi_2 <\xi_1|\hat\rho(t_0)|\xi_2> \\
    F(\xi)  \int\limits_{\eta_F(t_0)=\xi_1}^{\eta_F(t_1)
  =\xi} \Df \eta_F \int\limits_{\eta_B(t_0)=\xi_2}^{\eta_B(t_1)
 =\xi} \Df \eta_B\ e^{i S_K[\eta]}, 
 \label{t0t1} 
  \end{multline}
where $  S_K[\eta] \equiv S[\eta_F]-S[\eta_B]$ is the so-called Keldysh action and the integration goes along the Keldysh contour, see Fig.\ref{fig1}a.

In actual calculations it is convenient to rewrite (\ref{t0t1}) by extending temporal integration to infinity
by introducing an extended Keldysh contour,  see Fig.\ref{fig1}b. 
The convenience stems from the fact that all the dependence on $t_1$ now resides only in $F$ so that
\begin{multline}
  \langle F(\hat\varphi)\rangle_{t_1} =\int d\chi_1 \int d\xi_1 \int d\xi_2 <\xi_1|\hat\rho(t_0)|\xi_2> \\
   \int\limits_{\eta_F(t_0)=\xi_1}^{\eta_F(\infty) =\chi_1} \Df \eta_F 
  \int\limits_{\eta_B(t_0)=\xi_2}^{\eta_B(\infty)=\chi_1} \Df \eta_B  
   F\left(\frac{\eta_F(t_1)+\eta_B(t_1)}{2} \right)   \\
   \times e^{i S_K[\eta]}
    \label{t0t_infty_1}
\end{multline}
Let us now introduce new fields $\phi_c$ and $\phi_q$:
\begin{equation}
\phi_c = \frac{\eta_F+\eta_B}{2}, \qquad \phi_q = \eta_F -\eta_B.
\end{equation} 
The corresponding boundary conditions read
 \begin{gather}
\phi_c(t_0) = \frac{\xi_1+\xi_2}{2}, \qquad \phi_c(\infty) = \chi_1,\\
\phi_q(t_0) = \xi_1-\xi_2, \qquad \phi_q(\infty) = 0.
\end{gather}
Let us consider the scalar field theory with the lagrangian
\begin{equation}\label{lagr}
{\cal L} = \frac{1}{2} \dot\phi^2 - \frac{\lambda}{4 !} \phi^4 +J\phi .
\end{equation}
In terms of the new fields the action reads
\begin{multline}
S_K[\phi_c,\phi_q] = \int\limits_{t_0}^{\infty} dt [\dot\phi_c \dot\phi_q -\frac{\lambda}{4!}\phi_c\phi_q^3
-\frac{\lambda}{6}\phi_c^3\phi_q +J\phi_q]=\\
=\dot\phi_c(t_0) (\xi_1-\xi_2)- \int\limits_{t_0}^{\infty} dt\ 
( \phi_q A[\phi_c] +\frac{\lambda}{4!}\phi_c\phi_q^3),
\label{qc_action}
\end{multline} 
where
\begin{equation}
A[\phi_c]=\ddot\phi_c  +\frac{\lambda}{6}\phi_c^3 -J.
\end{equation}
We see that $A[\phi_c]=0$ corresponds to projecting onto the tree-level equation of motion for the lagrangian Eq.(\ref{lagr}). 

The systematic procedure we employ is expansion in $\phi_q$ in (\ref{qc_action}) around its saddle-point value. This expansion is, in fact, a quasiclassical one. This can be seen by restoring $\hbar$ in the action and replacing $\phi_q \to \hbar \phi_q$ so the only remaining dependence on $\hbar$  is in $\phi_q^3$ which is proportional to $\hbar^2$ and  
$$ e^{-i\hbar^2\frac{\lambda}{4!} \phi_c \phi_q^3 }\approx 1-i\hbar^2 \frac{\lambda}{4!}\phi_c \phi_q^3 +O(\hbar^4\phi_q^6) $$
and is built on top of the solution of the tree-level equations of motion $\ddot\phi_c^0  +\frac{\lambda}{6}(\phi_c^0)^3=0$. 
Explicitly \cite{initial1}:
\begin{equation}
\varphi^0_c(t)=\phi_{max} cn\left(\frac{1}{2};\sqrt{\frac{\lambda}{6}}\phi_{max} (t-t_0) + C \right),
 \label{class_sol}
\end{equation}
where $cn$ is the Jacobi elliptic function.
In what follows, in agreement with , we shall denote by LO the leading order contribution in $\phi_q$, etc.
(note the difference with notations in \cite{initial1,initial2}. 

\section*{lo approximation: analytical solution}

In the LO approximation we neglect the $\phi^3_q$ term in the Keldysh action. Integrating over $\phi_q$ in  (\ref{t0t_infty_1})
we get
\begin{multline}
 \langle F(\hat\varphi)\rangle_{t_1}^{LO} =  \int d\chi_1 \int d\xi_1 \int d\xi_2 <\xi_1|\hat\rho(t_0)|\xi_2>  \\
  \int\limits_{\phi_c(t_0)=\frac{\xi_1+\xi_2}{2}}^{\phi_c(\infty) =\chi_1}\Df \phi_c    \ F[\phi_c(t_1)] \\
  \int\frac{d\tilde p}{2\pi} e^{i\tilde p(\xi_1-\xi_2)}
  \delta(\tilde p-\dot\phi_c(t_0))\ \delta (A[\phi_c]).
  \label{almost_done}
\end{multline}
where we have introduced a new delta function to define "initial velocity" $\dot\phi_c(t_0) =\tilde p$. The initial value of $\phi_c^0$ is simply $\phi^0_c(t_0) = \frac{\xi_1+\xi_2}{2} \equiv \alpha$.  Denoting the corresponding classical solution by  $\phi_c^0$ we have $\ F[\phi_c(t_1)]=\ F[\phi_c^0(t_1)]$. Denoting  $ \xi_1-\xi_2=\beta$ and integrating over $\phi_c$ we get
\begin{gather}\label{LO}
\langle F(\hat\varphi)\rangle_{t_1}^{LO} = \int\frac{d\tilde p}{2\pi}  \int d\alpha f_W(\alpha,\tilde p,t_0)  F[\phi_c^0(t_1)] \\
f_W(\alpha,\tilde p,t_0) =\int d\beta <\alpha+\frac{\beta}{2}|\hat\rho(t_0)|\alpha-\frac{\beta}{2}> e^{i\tilde p\beta}, \nonumber 
\end{gather}
where $f_W(\alpha,\tilde p,t_0)$ is the Wigner function. 
We see that the LO approximation in $\phi_q$ corresponds to averaging over initial
conditions for classical trajectory with the weight given by the corresponding Wigner function.
Expression  (\ref{LO}) was earlier derived by different methods in \cite{initial0,initial1},
see also \cite{cosm} and \cite{chem1,chem2}.

For spatially inhomogeneous fields (\ref{LO}) is replaced by
\begin{multline}
\langle F(\hat\varphi)\rangle_{t_1}^{LO} = \\
\int D \tilde p({\bf x})  \int D\alpha({\bf x}) f_W[\alpha({\bf x}),\tilde p({\bf x}),t_0]  F[\phi_c^0(t_1,{\bf x})],
\label{LOI}
\end{multline}
where $\Df\phi(x)$ means the integration over 4-dimensional functions and symbol $D\phi({\bf x})$ -
over 3-dimensional ones and
\begin{equation}
\Box\phi_c^0  +\frac{\lambda}{6}(\phi_c^0)^3=0, \quad
\phi_c^0(t_0,{\bf x})=\alpha({\bf x}),\quad \dot\phi_c^0(t_0,{\bf x}) = \tilde p({\bf x}) \nonumber.
\end{equation}

\begin{figure}[h]
\includegraphics[width=0.5\textwidth]{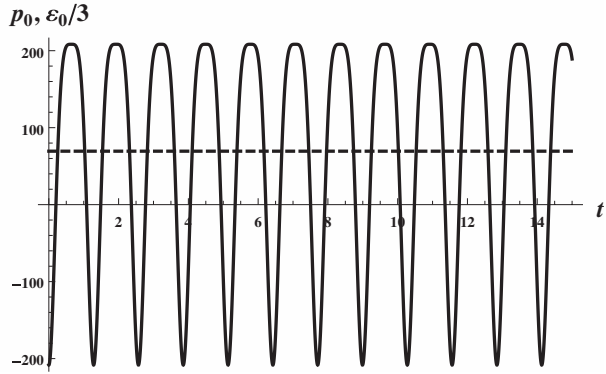}
\caption{Fig.2: Evolution of energy and pressure in the tree level approximation. The 
 parameter values are $p_0=1.5 \sqrt{2}$, $\alpha_0=1/p_0$, $A=10$, $\lambda=0.9$.}
 \label{ep_class}
\end{figure}

Let us now consider the evolution of the energy-momentum tensor
 \begin{gather}
  T^{\mu\nu}=\partial^{\mu}\varphi\partial^{\nu}\varphi-
  g^{\mu\nu}\left(\frac{1}{2}\partial_{\sigma}\varphi\partial^{\sigma}\varphi - 
  \frac{\lambda}{24}\varphi^4\right).
 \end{gather}
Of special interest here is dynamical interrelation between energy and pressure 
and possibility of reaching the "hydrodynamic" regime $\varepsilon = 3 p$.
In the case under consideration (homogeneous field) at the tree level 
\begin{gather}
\varepsilon_0=\frac{\dot\varphi^2}{2}+\frac{\lambda\varphi^4}{24}, \quad
p_0=\frac{\dot\varphi^2}{2}-\frac{\lambda\varphi^4}{24},
\end{gather}
where $\varphi=\varphi_0$ is the solution of the EoM eq.(\ref{class_sol}). The resulting dynamics of energy and pressure \cite{initial1} is shown in Fig.\ref{ep_class} from which one can see that there is no one-to-one relation between energy and pressure in this approximation.

Let us now turn to  of energy and pressure at the LO level. In ref. \cite{initial1} it was shown by numerical computation that averaging over initial conditions in (\ref{LO}) leads, after some transient period, to formation of well-defined equation of state. In this section we describe an analytical calculation supporting this conclusion. Following \cite{initial1}, let us use a Gaussian ansatz for the Wigner function  
\begin{gather}
f_W(\alpha,p,0)=\frac{1}{\alpha_0 p_0 \pi} e^{-\frac{(\alpha-A)^2}{\alpha_0^2}-\frac{p^2}{p_0^2}},
\end{gather}
where $A$ is the initial amplitude of the field and  $\alpha_0$ and $p_0$ are normalization constants. Let us make a change of variables $(\alpha_0,p_0) \to (\phi_{max},C)$ (see (\ref{class_sol})):
\begin{gather}
\int\frac{d\tilde p}{2\pi}  \int d\alpha \to \int |J|\ d \phi_{max} \ d C, \\
|J(\phi_{max})|=\sqrt{\frac{\lambda}{6}}\phi_{max}^2.
\end{gather}
In new variables the Wigner function reads
\begin{gather}\label{nwf}
f_W(\phi_{max},C,0)=\frac{1}{\alpha_0 p_0 \pi} e^{-\frac{(\phi_{max} cn\left(\frac{1}{2};C\right)-A)^2}{\alpha_0^2} }\\
\times e^{-\frac{\frac{\lambda}{6} \phi_{max}^4 sn\left(\frac{1}{2};C\right)^2 dn\left(\frac{1}{2};C\right)^2}{p_0^2}} .\nonumber
\end{gather}
Analytical integration over $\phi_{max}$ and $C$ is possible in the saddle point approximation, where
\begin{gather}
 f_W(\phi_{max},C,0) \approx \frac{1}{\alpha_0 p_0 \pi} e^{-\frac{(\phi_{max} -A)^2}{\alpha_0^2}
-\frac{C^2 A^4 \lambda }{6 p_0^2}}
\end{gather}
valid for $\alpha_0 \ll A$ and $p_0 \ll A^2 \sqrt{\lambda/6}$. Introducing a Fourier transform
\begin{gather}\label{um}
 cn\left(\frac{1}{2};\sqrt{\frac{\lambda}{6}}\phi_{max}t + C\right)
 =\sum_{k=-\infty}^{\infty} u_k e^{\frac{2 \pi i k}{T} \left(\sqrt{\frac{\lambda}{6}}\phi_{max} t +C\right)},\\
 u_m=\frac{1}{T} \int\limits_0^T cn\left(\frac{1}{2};t\right)e^{-i m t\frac{2 \pi}{T}} dt,\nonumber
\end{gather}
where $T=4 K(1/2)$ ( with $K(1/2)$ - the complete elliptic integral  of the first kind), 
we obtain the following general equation relating energy and pressure:
\begin{multline}\label{eos}
 p_{\rm LO} = \varepsilon_{\rm LO} \left(-8 \left(\frac{2 \pi}{T}\right)^2 \sum_{k=-\infty}^{\infty} \right.
 \sum_{l=0}^{\infty} k\ l\  u_k u_l 
  e^{-\frac{6 \pi^2 p_0^2}{\lambda A^4 T^2}(k+l)^2}  \\
 \times \left. e^{-\frac{\alpha_0^2 \pi^2 \lambda }{6 T^2}(k+l)^2 t^2}
 \mbox{cos}\left(\frac{2 \pi  A (k+l)}{T}\sqrt{\frac{\lambda}{6}}t\right)
 -1\right),
\end{multline}
where $\varepsilon_{\rm LO} = \lambda A^2 / 24$.

Let us consider the large time limit $t \to \infty$. The resulting expressions are conveniently written using the sum
\begin{gather}
I(q)= -\left(\frac{2 \pi}{T}\right)^2 \sum\limits_{k=-\infty}^{\infty} k(q-k) u_k u_{q-k}=\\
 \frac{1}{T}\int\limits_{0}^{T} \left (\frac {d\ cn \left(\frac{1}{2}; t \right)}{dt}\right)^2 
 e^{ -\frac{2 \pi i}{T}  q t}, \nonumber
\end{gather} 
where $q$ is an integer number. Using the properties of the coefficients $u_k$ in (\ref{um}) it is easy to prove that $I(q)=I(-q)$.
The leading asymptotic at $t \to \infty$ comes from the term with $q=0$. The corresponding sum can be calculated analytically, $I(0)=1/3$, so that 
\begin{equation}\label{eos0}
 p_{\rm LO} (t\to\infty) = \varepsilon_{LO} (4 I(0) -1)= \frac{\varepsilon_{LO}}{3}.
\end{equation}
We see that indeed in the limit $t \to \infty$ we recover the ultrarelativistic equation of state $\varepsilon = 3 p$. The leading correction to (\ref{eos0}) comes form the term with $q= \pm 2$. Explicitly:
\begin{multline}\label{eos1}
 p_{LO}(t\to\infty) =\varepsilon_{LO}\left[ \frac{1}{3} +\right.\\
 \left. 8 I(2) e^{-\frac{24 \pi^2 p_0^2}{\lambda A^4 T^2}}  
  e^{-\frac{2 \alpha_0^2 \pi^2 \lambda }{3 T^2} t^2}
 \mbox{cos}\left(\frac{4 \pi  A }{T}\sqrt{\frac{\lambda}{6}}t\right)+ ... \right],
\end{multline}
where $I(2)\approx -0.12$.  From (\ref{eos1}) we see that "thermalization time" $t_{\rm th}$ can be estimated as
\begin{equation}
t_{\rm th} \sim \sqrt{\frac{3}{2}} \; \frac{T}{\pi \alpha_0 \sqrt{\lambda}}.
\end{equation}
In Fig. 3 we compare numerical results for $p_{\rm LO}$ and $\varepsilon_{LO}$ with analytical expression (\ref{eos}) in which terms up to $q=6$ were retained
($I(4)\approx -0.04,\ I(6)\approx -0.006$).  
\begin{figure}[h]
 \includegraphics[width=0.5\textwidth]{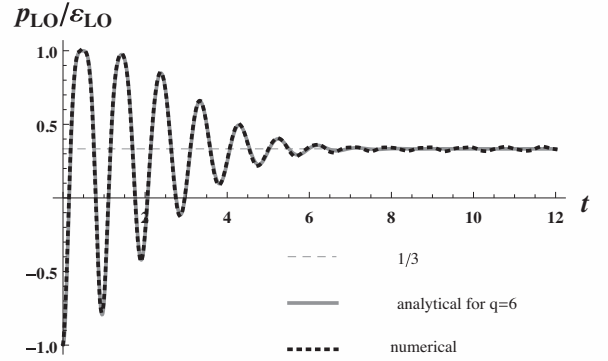}
 \caption{Fig. 3: Pressure relaxation: comparison of numerical result an analytical expression (\ref{eos}) with terms up to $q=6$ taken into account. The 
 parameter values are $p_0=1.5 \sqrt{2}$, $\alpha_0=1/p_0$, $A=10$, $\lambda=0.9$.}
\end{figure}
We see that agreement between numerical and analytical results is very good.

\section*{NLO corrections}
 Let us now consider NLO corrections.  Their importance is not only in a possibility of obtaining more accurate 
 expressions for the above-considered observables but, importantly, in opening the way for calculation of various correlation functions describing, in particular, transport properties of the system. Technically we should return to 
 eq.(\ref{qc_action}) and expand $\phi_q^3$ term in the action:
 \begin{multline}
 e^{-i\frac{\lambda}{4!}\int\limits_{t_0}^{\infty} dt'\phi_c\phi_q^3}\approx 
 1- \frac{i\lambda}{4!}\int\limits_{t_0}^{\infty} dt'\phi_c(t')\phi_q^3(t') +O(\phi_q^6)
 \end{multline}
 Using procedure described above and relations
\begin{equation}
\frac{\delta }{\delta J(t) } e^{i S_K[\phi_c,\phi_q]} = i \phi_q(t) e^{i S_K[\phi_c,\phi_q]}
\label{variation}
\end{equation}
and 
\begin{equation}
\frac{\delta \phi_c^0(t_1)}{\delta J(t')}  = 0\ \mbox{ if }\ t'\geq t_1,
\label{greate}
\end{equation}
which follows from causality,
one can obtain for NLO correction
\begin{multline}
\langle  F(\hat\varphi)\rangle_{t_1}^{\rm NLO} = 
\int\frac{d\tilde p}{2\pi}\int d\alpha\ f_W(\alpha,\tilde p,t_0)\\
\times\left(F[\phi_c^0(t_1)]+
\left. \frac{\lambda}{4!}\int\limits_{t_0}^{t_1} dt' \phi_c^0(t') \frac{\delta^3 F[\phi_c^0(t_1)]}{\delta J^3(t')}
 \right|_{J=0} \right).
 \label{deltaF_1}
\end{multline}
For simplicity let us denote solution of the classical EoM  $\phi_c^0$ as $\phi$ and variations of this solution over source as:
\begin{gather}
\frac{\delta \phi(t_1)}{\delta J(t')} = \Phi_1(t_1,t'), \ 
\frac{\delta^2 \phi(t_1)}{\delta J^2(t')} = \Phi_2(t_1,t'),\\ 
\frac{\delta^3 \phi(t_1)}{\delta J^3(t')} = \Phi_3(t_1,t').
\end{gather}
Then 
\begin{multline}
\frac{\delta^3 F[\phi(t_1)] }{\delta J^3(t')}= \frac{dF}{d\phi}\Phi_3(t_1,t') +
3\frac{d^2F}{d\phi^2} \Phi_2(t_1,t')\Phi_1(t_1,t')\\
+\frac{d^3 F}{d\phi^3}\Phi_1(t_1,t')^3 .
\end{multline}
We can find variations of the field $\phi$ with help of equation of motion as
\begin{equation}
\frac{\delta^3}{\delta J^3(t')} (\ddot\phi + \frac{\lambda}{6}\phi^3 -J)_{t_1} =0.
\end{equation}
In such a way we obtain the set of the differential equations on the variations:
\begin{gather}
\hat L_{t_1} \Phi_1(t_1,t') = \delta(t_1-t'), \label{var1}\\
\hat L_{t_1} \Phi_2(t_1,t') = -\lambda \phi(t_1)\Phi_1^2(t_1,t'),\nonumber\\
\hat L_{t_1} \Phi_3(t_1,t') =  -\lambda \Phi_1^3(t_1,t') 
- 3 \lambda \phi(t_1) \Phi_1(t_1,t')\Phi_2(t_1,t'),\nonumber\\
\hat L_t=\partial^2_t + \frac{\lambda}{2} \phi^2(t).\nonumber
\end{gather}
As  we can solve them for explicit EoM 
we derive the answer for NLO correction eq.(\ref{deltaF_1}).

In similar way we can obtain expressions for 2-point
 correlation functions in the Keldysh formalism:
 \begin{gather}
<\phi_c(t_1)\phi_c(t_2)>=
\int\frac{d\tilde p}{2\pi}  \int d\alpha f_W(\alpha,\tilde p,t_0) \phi_c^0(t_1)\phi_c^0(t_2),\nonumber\\
<\phi_c(t_1)\phi_q(t_2)>=
-i \int\frac{d\tilde p}{2\pi}  \int d\alpha f_W(\alpha,\tilde p,t_0)
\frac{\delta\phi_c^0(t_1)}{\delta J(t_2)}\nonumber\\
=-i \int\frac{d\tilde p}{2\pi}  \int d\alpha f_W(\alpha,\tilde p,t_0)
\Phi_1(t_1,t_2),\nonumber\\
<\phi_q(t_1)\phi_q(t_2)>=0 \mbox{ by construction.}
\end{gather}

It is interesting  that the first variation $\Phi(t,t')$ can be expressed in the term of the 
Jacobi elliptical functions. One can note that 
\begin{multline}
\partial_t\ [\ddot\phi_c^0(t) + \frac{\lambda}{6}(\phi_c^0(t))^3]=0\\
\mbox{gives}\\
[\partial^2_t +\frac{\lambda}{2} (\phi_c^0(t))^2]\dot\phi_c^0(t)=\hat L_{t} \dot\phi_c^0(t) =0.
\end{multline}
It means that $\dot\phi_c^0(t)\equiv f_1(t) $ is the first particular solution of eq. ( \ref{var1})
on Green's function $G(t,t')$
\begin{equation}
\Phi_1(t,t') =  G(t,t') = \theta(t-t')[c_1(t') f_1(t) + c_2(t') f_2(t)]. 
\label{green}
\end{equation}
By usual procedure we can construct the second particular 
solution and obtain expression for the first variation
\begin{multline}
 \Phi_1(t_1,t_2)=\theta(t_1-t_2)\frac{6}{\lambda \phi_{max}^2}\\
 \times \left[cn\left(\frac{1}{2};\sqrt{\frac{\lambda}{6}}\phi_{max}t_1 +C\right)\right.
 \dot cn\left(\frac{1}{2};\sqrt{\frac{\lambda}{6}}\phi_{max}t_2 +C\right)-\\
 cn\left(\frac{1}{2};\sqrt{\frac{\lambda}{6}}\phi_{max}t_2 +C\right)
 \dot cn\left(\frac{1}{2};\sqrt{\frac{\lambda}{6}}\phi_{max}t_1 +C\right)+\\
\dot cn\left(\frac{1}{2};\sqrt{\frac{\lambda}{6}}\phi_{max}t_1 +C\right)
\left.\dot cn\left(\frac{1}{2};\sqrt{\frac{\lambda}{6}}\phi_{max}t_2 +C\right)(t_1-t_2)\right],
\end{multline}
where
\begin{equation}
\frac{d\ cn(k^2;t)}{dt}=-sn(k^2;t)\cdot dn(k^2;t).
\end{equation}

\section*{conclusions}
Let formulate once again the main results obtained in the paper:
\begin{enumerate}
\item The systematic procedure of computing quantum corrections in the framework of Keldysh formalism is described.
\item Analytical expressions for pressure relaxation in the scalar field model of \cite{initial1} are presented.
\item Explicit equations for the next-to-leading order corrections are written down.
\end{enumerate}

The authors are grateful to A.G. Semenov for many useful discussions of Keldysh formalism and its applications.

\end{document}